# An Improved Integer Multiplicative Inverse (modulo $2^w$)

Jeffrey Hurchalla
jeffrey@jeffhurchalla.com

April 8, 2022

**Abstract**

This paper presents an algorithm for the integer multiplicative inverse (mod $2^w$) which completes in the fewest cycles known for modern microprocessors, when using the native bit width $w$ for the modulus $2^w$. The algorithm is a modification of a method by Dumas, and for computers it slightly increases generality and efficiency. A proof is given, and the algorithm is shown to be closely related to the better known Newton's method algorithm for the inverse. Simple direct formulas, which are needed by this algorithm and by Newton's method, are reviewed and proven for the integer inverse modulo $2^k$ with $k$ = 1,2,3,4, or 5, providing the first proof of the preferred formula with $k$=4 or 5.

## 1  Introduction

The integer multiplicative inverse (modulo $2^w$) of an odd integer $a$ is the integer $x$ that satisfies

$$ax \equiv 1 \pmod{2^w} \tag{1}$$

where $w$ is typically set to the bit width of arithmetic operations on a computer.

We will refer to $x$ as simply the "multiplicative inverse" or "inverse". The inverse $x$ is a unique value (mod $2^w$) for any odd integer $a$, and it does not exist for any even integer $a$.

Applications for the inverse include Montgomery multiplication [1][16], computing exact division by an integer constant [2], testing for a zero remainder after dividing by a constant [3], and integer division [4].

The inverse can be calculated in at least three different ways. The first and most well known method is the extended Euclidean algorithm [5]. The second method is inspired by Newton's method for approximating a reciprocal [6], and we follow a convention of simply calling it "Newton's method"[1] [10]. The third method was discovered by Dumas [7], and is expanded upon in this paper for implementation on a computer.

For background, calculating the inverse via Newton's method is almost always more efficient on a computer than via the extended Euclidean algorithm. Given the modulus $2^w$, Newton's method has order of complexity $O(\log w)$, as compared to $O(w)$ for the extended Euclidean. Newton's method also has a better constant factor in its expected running time due to the fact that it uses no divisions. Granlund and Montgomery [9] first mentioned Newton's method for the integer inverse, Warren [10] provided an easy to follow algorithm and proof for it, and academic papers [7][11][12] and blog entries [8][13][14] further explored it.

---
1   More accurately it is Newton's method over p-adic numbers, or a use of Hensel lifting.

Given the modulus $2^w$, Dumas' Algorithm 3 [7] has nearly the same number and type of operations as Newton's method. In practice however, Dumas' algorithm can complete faster on modern CPUs because a CPU can exploit instruction level parallelism when executing it [8]. Section 6 shows its improved latency compared to Newton's method. The algorithm in this paper further improves Dumas' algorithm for use on a computer.

## 2 Computing the Multiplicative Inverse

We present a different derivation of Dumas' Algorithm 3 that provides improvements to its flexibility and efficiency on a computer.

For the sake of readability, we use the expression *pow(2, n)* to represent an integer $2^n$. For example, $y^{pow(2,n)}$ would represent $y^{2^n}$.

Let *a* be an odd integer for which we wish to find the multiplicative inverse (mod $2^w$). Let *w* be a positive integer. In practice we set *w* to the bit width of the arithmetic we will use on a computer; most commonly *w* is 32 or 64, for a 32 bit or 64 bit computer architecture. Let *k* be a positive integer such that *k* divides *w*, and such that (*w*/*k*) is a power of 2. We usually wish to use *k*<=5 (in order to use simple direct formulas from section 5), so if a *k* in that range does not exist, increase *w* by the minimal amount needed to obtain a satisfactory *k*; the algorithm's result will suffice for your original value of *w*. Common choices for *k* would be 1, 2, or 4. We can note in advance that doubling the size of *k* decreases the number of iterations required by the upcoming recurrence (4) by 1.

Set $x_0$ to any integer that satisfies
$$ax_0 \equiv 1 \ (mod \ 2^k) \tag{2}$$
Section 5 lists formulas for $x_0$ for values of *k*=1, 2, or 4. For example, if we are using *k*=1, then $x_0$=1 satisfies (2).

Set *y* to any integer that satisfies
$$y \equiv 1 - ax_0 \ (mod \ 2^w) \tag{3}$$

Define the discrete recurrence relation
$$x_{n+1} \equiv x_n(1 + y^{pow(2,n)}) \ (mod \ 2^w) \tag{4}$$

Set *p* to the integer
$$p = \log_2(w/k) \tag{5}$$
And perform *p* iterations of recurrence relation (4).
The resulting $x_p$ is the desired integer inverse of *a*, satisfying
$$ax_p \equiv 1 \ (mod\, 2^w) \tag{6}$$

The following C function implements the algorithm, using *w*=64, *k*=4, and thus *p*=4:

```
1.   uint64_t integer_inverse(uint64_t a)
2.   {
3.     assert(a%2 == 1);
4.     uint64_t x0 = (3*a)^2;     // See section 5, formula 3.
5.     uint64_t y = 1 - a*x0;
6.     uint64_t x1 = x0*(1 + y);
7.     y *= y;
8.     uint64_t x2 = x1*(1 + y);
9.     y *= y;
10.    uint64_t x3 = x2*(1 + y);
11.    y *= y;
12.    uint64_t x4 = x3*(1 + y);
13.    return x4;
14.  }
```

*Figure 1: C code for the 64 bit integer inverse.*

Note that any unsigned computer arithmetic instruction that has bit width *w* automatically gets performed modulo $2^w$. This is reflected in C and C++; the data types uint32_t and uint64_t respectively ensure arithmetic modulo $2^{32}$ and $2^{64}$.

## 3  Proof

Rearranging (3),
$$ax_0 \equiv 1-y \equiv 1-y^{pow(2,\,0)} \ (mod\, 2^w) \tag{7}$$
Assume for an arbitrary value of integer *n* that
$$ax_n \equiv 1 - y^{pow(2,\,n)} \ (mod\, 2^w) \tag{8}$$
Multiply both sides of (4) by *a*.
$$ax_{n+1} \equiv ax_n(1 + y^{pow(2,\,n)}) \ (mod\, 2^w) \tag{9}$$
Substitute (8) for $ax_n$.
$$ax_{n+1} \equiv (1 - y^{pow(2,\,n)})(1 + y^{pow(2,\,n)}) \equiv 1 - y^{pow(2,\,n+1)} \ (mod\, 2^w) \tag{10}$$
By induction, base case (7) and steps (8) (10) prove that for all n >= 0,
$$ax_n \equiv 1 - y^{pow(2,\,n)} \ (mod\, 2^w) \tag{11}$$

Substitute (3) for $y$.
$$ax_n \equiv 1-(1-ax_0)^{pow(2,\,n)} \pmod{2^w} \tag{12}$$
By definition of congruence, there exists an integer $q$ such that we can restate (2) as
$$ax_0 = 1 - q2^k \tag{13}$$
Substitute (13) into (12).
$$ax_n \equiv 1-(1-(1-q2^k))^{pow(2,\,n)} \equiv 1 - q^{pow(2,\,n)} 2^{k\,pow(2,\,n)} \pmod{2^w} \tag{14}$$
Since (5) sets $p = \log_2(w/k)$,
$$ax_p \equiv 1 - q^{pow(2,\,p)} 2^{k\,pow(2,\,\log_2(w/k))} \equiv 1 - q^{pow(2,\,p)} 2^w \equiv 1 \pmod{2^w} \tag{15}$$

Thus $x_p$ is the integer multiplicative inverse of $a$ (mod $2^w$).

## 4 Relationship to Newton's Method

We can show that the algorithm of Section 2 is closely related to Newton's method.

Restating recurrence (4),
$$\begin{aligned} x_{n+1} &\equiv x_n(1 + y^{pow(2,\,n)}) \pmod{2^w} \\ x_{n+1} &\equiv x_n(2 - (1 - y^{pow(2,\,n)})) \pmod{2^w} \end{aligned} \tag{16}$$
Substitute (11) into (16).
$$x_{n+1} \equiv x_n(2 - ax_n) \pmod{2^w} \tag{17}$$
The recurrence for the multiplicative inverse via Newton's method [10] is given as
$$x_{n+1} = x_n(2 - ax_n) \tag{18}$$

Taking this modulo $2^w$ results in (17), and thus for all $n$, the Newton's method recurrence and Section 2's recurrence (4) produce equivalent values modulo $2^w$ (i.e. up to $w$ bits).

## 5 Formulas to Set $x_0$, with Proofs

Simply restating (2), we wish to find an integer $x_0$ such that
$$ax_0 \equiv 1 \pmod{2^k} \tag{19}$$

**Formula 1 ($k=1$)**
For $k=1$, set $x_0=1$.
**Proof:**
Since $a$ is an odd integer, $a*1 \equiv 1 \pmod{2^1}$, and by substitution $ax_0 \equiv 1 \pmod{2^k}$, which is congruence (19).

**Formula 2 ($k$=2)**
For $k$=2, set $x_0$=$a$.
**Proof:**
Let the integer $q$=$a$/4, and the integer $r$=$a$%4. Thus
$$\begin{aligned} a &= 4q+r \\ a &\equiv r \ (mod\ 4) \\ a^2 &\equiv r^2 \ (mod\ 4) \end{aligned} \tag{20}$$
Since $a$ is odd, $r$ belongs to the set {1, 3}. By testing every element of the set we see that
$$r^2 \equiv 1 \ (mod\ 4) \tag{21}$$
Thus by (20),
$$a^2 \equiv r^2 \equiv 1 \ (mod\ 4) \tag{22}$$
Since $a*a \equiv 1 \ (mod\ 2^2)$, by substitution $ax_0 \equiv 1 \ (mod\ 2^k)$.
This same approach can also show that for $k$=3, $x_0$=$a$ always satisfies (19). We focus on $k$=2 because $k$=3 is rarely needed.

**Formula 3 ($k$=4 and $w$>=4)**
For $k$=4, set $x_0$=**XOR((3$a$)%2$^w$, 2)**. The formula is attributed to Peter Montgomery [4].
**Proof:**
Let the integer $q$=$a$/16, and the integer $r$=$a$%16. Thus
$$\begin{aligned} a &= 16q+r \\ a &\equiv r \ (mod\ 16) \end{aligned} \tag{23}$$
Since $a$ is odd, $r$ belongs to the set {1, 3, 5, 7, 9, 11, 13, 15}. By testing every element of the set we see that
$$r * XOR(3r,2) \equiv 1 \ (mod\ 16) \tag{24}$$
Using (23) with (24),
$$a * XOR(3r,2) \equiv 1 \ (mod\ 16) \tag{25}$$
By (23) and the fact that $w$>=4,
$$\begin{aligned} 3a &\equiv 3r \ (mod\ 16) \\ (3a)\%2^w &\equiv 3r \ (mod\ 16) \end{aligned} \tag{26}$$
Thus, the result of $3a$ % $2^w$ must have the same value in its second bit as the result of $3r$. We can observe that XORing an integer with 2 flips the second bit of the integer and leaves the rest of the integer unchanged. Since the results $3a$ % $2^w$ and $3r$ both contain the same value at their second bits, both results will continue to have equal second bits after being XORed with 2, and the rest of their bits will be unchanged. Continuing from (26),
$$XOR((3a)\%2^w, 2) \equiv XOR(3r, 2) \ (mod\ 16) \tag{27}$$
Substituting (27) into (25),
$$a * XOR((3a)\%2^w, 2) \equiv 1 \ (mod\ 16) \tag{28}$$
By substitution we can see that $a*x_0 \equiv 1 \ (mod\ 2^k)$, which is congruence (19).
This same approach can also show that for $k$=5, $x_0$=XOR((3$a$) % 2$^w$, 2) always satisfies (19).

**Formula 4 (**an alternate formula for **k=4** and **w>=4)**
For *k*=4, set $x_0 =(XOR(a, 2) - ((a+a)\%2^w))\%2^w$.
This formula was found by brute force testing combinations of operators and integer constants across candidate formulas $x_0$= (*r* **op1** *constant1*) **op2** (*r* **op3** *constant2*), where *r* is a variable integer, and **op1**, **op2**, and **op3** were any of the operators XOR, AND, OR, add, subtract, leftshift; a candidate formula for $x_0$ was successful if it satisfied $rx_0 \equiv 1 \ (mod\ 16)$ for all odd values *r* from 1 to 15.
Note that Formula 4 usually requires one more instruction to implement than Formula 3. An exception to this is usually SIMD implementations.
**Proof:**
Let the integer *q*=*a*/16, and the integer *r*=*a*%16. Thus
$$a = 16q + r$$
$$a \equiv r \ (mod\ 16) \tag{29}$$
Since *a* is odd, *r* belongs to the set {1, 3, 5, 7, 9, 11, 13, 15}. By testing every element of the set we see that
$$r * (XOR(r,2) - (r+r)) \equiv 1 \ (mod\ 16) \tag{30}$$
Using (29) with (30),
$$a * (XOR(r,2) - (a+a)) \equiv 1 \ (mod\ 16) \tag{31}$$
By (29), $a \equiv r \ (mod\ 16)$. Thus *a* must have the same value in its second bit as *r*. We can observe that XORing an integer with 2 flips the second bit of the integer and leaves the rest of the integer unchanged. Since *a* and *r* contain the same value at the second bit, both integers will continue to have equal second bits after being XORed with 2, and the rest of their bits will be unchanged. Thus $a \equiv r \ (mod\ 16)$ implies that
$$XOR(a,2) \equiv XOR(r,2) \ (mod\ 16) \tag{32}$$
Substituting (32) into (31),
$$a * (XOR(a,2) - (a+a)) \equiv 1 \ (mod\ 16) \tag{33}$$
And since *w*>=4,
$$a * (XOR(a,2) - (a+a)\%2^w)\%2^w \equiv 1 \ (mod\ 16) \tag{34}$$
By substitution we can see that $a * x_0 \equiv 1 \ (mod\ 2^k)$, which is congruence (19).

# 6  Efficiency

We compare in this section the expected number of CPU cycles to complete a 64 bit inverse (i.e. modulo $2^w$ with *w* = 64), based on a performance model. We will also briefly consider throughput at the end of this section. For our performance model, we specify that an addition (or subtraction) requires 1 cycle to complete and that a multiplication requires 3 cycles to complete, and that multiplications are pipelined and that an addition can issue and execute in parallel with a multiplication. This model also specifies that the operation 3*a* can be completed in one cycle (via an add with shift instruction, or the x86 LEA instruction), and that an XOR operation requires one cycle. This model roughly corresponds to x86 Intel CPU non-SIMD instruction timings from the last 12 years [15], and it reflects typical capabilities of modern CPUs.

Let's implement Newton's method [10] for the 64 bit inverse:

```
15.    uint64_t newtons_inverse(uint64_t a)
16.    {
17.       assert(a%2 == 1);
18.       uint64_t x0 = (3*a)^2;      // See section 5, formula 3.
19.       uint64_t x1 = x0*(2 - a*x0);
20.       uint64_t x2 = x1*(2 - a*x1);
21.       uint64_t x3 = x2*(2 - a*x2);
22.       uint64_t x4 = x3*(2 - a*x3);
23.       return x4;
24.    }
```
*Figure 2: C code for the 64 bit inverse using Newton's method.*

Inspired by the example of Reynolds [8], let us consider the cycle timings of Figure 2 under the performance model we described. We will start on cycle 0 with $3*a$. Operations are listed in order of the cycle on which they can issue. If more than one operation could issue on the same cycle, they would be listed on the same row, though we will see this never happens with Figure 2. Since every instruction in Figure 2 depends upon its preceding instruction, there is one long dependency chain throughout the Newton's method algorithm:

```
Cycle number       Operation(s)
0.                 tmp = 3*a         (usually a single cycle operation)
1.                 x0 = tmp ^ 2      (^ is xor)
2.                 tmp = a*x0
3.                                   (no instruction possible, waiting on tmp)
4.                                   (no instruction possible, waiting on tmp)
5.                 tmp = 2 – tmp
6.                 x1 = x0 * tmp
7.
8.
9.                 tmp = a*x1
10.
11.
12.                tmp = 2 – tmp
13.                x2 = x1 * tmp
14.
15.
16.                tmp = a*x2
17.
18.
19.                tmp = 2 – tmp
20.                x3 = x2 * tmp
21.
22.
23.                tmp = a*x3
24.
```

```
25.
26.                     tmp = 2 − tmp
27.                     x4 = x5 * tmp
28.
29.
30.                     return x4
```

Let's implement Dumas' original Algorithm 3 [7] without looping for the 64 bit inverse:

```
25.     uint64_t original_dumas_inverse(uint64_t a)
26.     {
27.        assert(a%2 == 1);
28.        uint64_t y = a - 1;
29.        uint64_t u0 = 2 - a;
30.        y *= y;
31.        uint64_t u1 = u0*(1 + y);
32.        y *= y;
33.        uint64_t u2 = u1*(1 + y);
34.        y *= y;
35.        uint64_t u3 = u2*(1 + y);
36.        y *= y;
37.        uint64_t u4 = u3*(1 + y);
38.        y *= y;
39.        uint64_t u5 = u4*(1 + y);
40.        return u5;
41.     }
```
*Figure 3: C code for the 64 bit inverse using Dumas' algorithm 3.*

Let's consider the cycle timings for Figure 3, starting on cycle 0 with its subtraction a - 1. Operations are listed in order of the cycle on which they can issue, and when more than one operation can issue on the same cycle, they are listed on the same row:

```
Cycle number       Operation(s)
0.                 y = a − 1
1.                 u0 = 2 − a,          y *= y
2.                 (no instruction possible, waiting on y)
3.                 (no instruction possible, waiting on y)
4.                 tmp = (1 + y),       y *= y
5.                 u1 = u0 * tmp,
6.
7.                 tmp = (1 + y),       y *= y
8.                 u2 = u1 * tmp
9.
10.                tmp = (1 + y),       y *= y
11.                u3 = u2 * tmp
12.
13.                tmp = (1 + y),       y *= y
14.                u4 = u3 * tmp
15.
16.                tmp = (1 + y)
17.                u5 = u4 * tmp
18.
```

```
19.
20.                    return u5
```

There are two dependency chains (one for *y* and one for *u*) executing mostly in parallel. As a result Dumas' algorithm completes in 20 cycles, compared to 30 cycles for Newton's method.

And finally let's consider the timings for our earlier Figure 1, which implemented the 64 bit inverse using the algorithm presented in this paper:

```
Cycle number       Operation(s)
0.                 tmp = 3*a        (usually a single cycle op)
1.                 x0 = tmp ^ 2     (^ is xor)
2.                 tmp = a*x0
3.
4.
5.                 y = 1 − tmp
6.                 tmp = (1 + y),      y *= y
7.                 x1 = x0 * tmp
8.
9.                 tmp = (1 + y),      y *= y
10.                x2 = x1 * tmp
11.
12.                tmp = (1 + y),      y *= y
13.                x3 = x2 * tmp
14.
15.                tmp = (1 + y)
16.                x4 = x3 * tmp
17.
18.
19.                return x4
```

Compared to Figure 3's timings, Figure 1 completes one cycle faster, mostly due to its ability to use the efficient formula 3 (or formula 4) from section 5. Under our performance model of a typical modern microprocessor, this paper's algorithm (Figure 1) improves expected latency of the 64 bit inverse by ~60% over Newton's method (Figure 2) and by ~5% over Dumas' Algorithm 3 as presented in his paper (Figure 3).

We will briefly mention throughput, which unfortunately needs specific context to analyze. It depends not only on these functions, but largely upon surrounding code, data dependencies, and the CPU micro-architecture. Nonetheless, we may offer two cautious observations: If there is a loop-carried dependency on the result of the inverse, the lowest latency functions for the inverse might tend to help throughput. If a loop contains the inverse and has a large number of multiply operations, the combined pressure on the CPU's multiplication execution unit(s) may hinder throughput; in such a case, an inverse function with a low number of multiplies might help (this paper's function and Newton's method have the lowest number of multiplies).

# 8  Conclusion

Dumas' algorithm for the inverse has a strong advantage over Newton's method due to its ability to exploit instruction level parallelism (from pipelining and/or superscalar execution units) available in modern microprocessors. The algorithm in this paper expands upon Dumas' work to allow for a more efficient starting value for its recurrence sequence, typically using Formula 3 from Section 5. This results in the lowest number of cycles currently known for calculating the multiplicative inverse of an integer (modulo $2^w$) on a computer, when $w$ is less than or equal to the native bit width of a CPU's arithmetic instructions.